
\input harvmac

\Title{\vbox{\hbox{HUTP--95/A004,}}}
{\vbox{\centerline{Topological Reduction of 4D SYM to 2D
$\sigma$--Models}}}
\vskip .2in
\centerline{M. Bershadsky, A. Johansen, V. Sadov and C. Vafa}
\vskip .2in
\centerline{Lyman Laboratory of Physics, Harvard
University}
\centerline{Cambridge, MA 02138, USA}
\vskip .3in
By considering a (partial) topological twisting of
supersymmetric Yang-Mills compactified on a 2d space with
`t Hooft magnetic flux turned on
we obtain a supersymmetric
$\sigma$-model in 2 dimensions.  For $N=2$ SYM this
maps Donaldson observables on products
of two Riemann surfaces to quantum cohomology ring of moduli space
of flat connections on a Riemann surface.
For $N=4$ SYM it maps $S$-duality to $T$-duality
for $\sigma$-models on moduli space of solutions to Hitchin
equations.

\Date{1/95}

\newsec{Introduction}

One of the main sources of insights into the dynamics
of 4 dimensional quantum
field theories comes from analogies with simpler 2 dimensional
quantum field theories.  It is the aim of this paper to make
this analogy more precise in the context of supersymmetric
gauge theories in 4 dimensions and special
classes of supersymmetric $\sigma$-models in 2 dimensions.  This
will
in particular allow us to map Donaldson observables on
products of two Riemann surfaces to quantum cohomology ring
of the moduli
space
of flat connections on one of the surfaces.  In the context of
$N=4$ YM, this reduction allows us to map $S$-duality
to $T$-duality of certain $\sigma$-models, thus relating
electric-magnetic
duality
to momentum-winding duality of $\sigma$-models.

The basic idea is rather simple.  We consider a Euclidean quantum
field theory on a product of two Riemann surfaces $\Sigma \times
C$ in the limit where the size of  one of them, say $C$ shrinks
to zero.  This gives rise to a quantum field theory on $\Sigma$.
  The reduction of 4d Yang-Mill theory to
2d is in general very complicated due to the fact that different
regimes
of field configurations of the 4d theory result in different 2d
effective
theories which are related to each other in a complicated way.
For four dimensional gauge theories a single regime of field
configuration
can be singled out
by restricting attention to the sectors of path integral with
non-trivial `t Hooft magnetic flux on $C$
(which thus avoid having reducible gauge connections).

Starting from  supersymmetric quantum field theories in 4d,
we expect to get a supersymmetric theory in 2d.  This is
only the case when $C$ is a torus with periodic
boundary conditions.
In the case $C$ is a torus, turning on the flux unfortunately
leads to a trivial quantum field theory on $\Sigma$ with
all the degrees of freedom frozen out.  However one can
consider a topologically twisted version of the 4d theory, which
does give rise to a non-trivial
supersymmetric 2d theory for any choice of
$C$ with genus greater than 1.  In fact we can consider a fully
twisted
topological theory
in 4 dimensions giving rise to a topological $\sigma$-model in 2
dimensions
or we can consider partial twisting of the 4 dimensional theory
only along the $C$ directions and obtain an untwisted
supersymmetric $\sigma$-model
on $\Sigma$.  Each twisting has its virtue:  The fully twisted
version is useful in that the topological amplitudes in 4d,
being independent of the size of $C$, are directly related to
topological amplitudes of the $\sigma$-model in 2d.  The partially
twisted theory, on the other hand, even though it depends on the
size of $C$,
 carries more information about
non-topological aspects of the 4d theory
\foot{A possible relation between the dynamics of the 4d
supersymmetric
theories and those of corresponding $\sigma$-models has been
conjectured in
ref.\ref\joh{A.Johansen, Infinite conformal algebras in
supersymmetric theories
on four manifolds. Preprint NBI-HE-94-34}.}.  We will consider both
twistings
in this paper.

Let us first consider $N=1$ supersymmetric theory.
The manifold $M^4 $ has a product structure and  therefore the
holonomy group is reduced to
$U(1)_\Sigma \times U(1)_C$, where
each $U(1)$ is the holonomy of the corresponding surface.  The
$U(1)$ charges of the supersymmetry generator is given
by $(\pm {1\over 2},\pm {1\over 2})$.  In addition
the supercharge carries an $R$ charge $\pm 1 $
(which with an appropriate choice
of $N=1$ theories with matter is anomaly free) which is correlated
with the chirality of the spinor (even or odd number of minus signs
in their $U(1)_\Sigma \times U(1)_C$ charge).  If we twist the
$U(1)_C$ by
adding
${-R/2}$ to it, we find that there are two components of the
supersymmetry which become spin $0$ in the $C$ direction and are
both of the form $(+{1\over 2},0)$.  We thus end up with a $(2,0)$
supersymmetric theory on $\Sigma$ for arbitrary choice of $C$
with genus greater than 1.
If we had in addition twisted the $U(1)_\Sigma$ by adding
${-R/2}$ we would have obtained a topologically twisted
$(2,0)$ theory in 2d.  With standard twistings,
in the case of $N=2$ theories the same construction leads
to a  $(2,2)$, and for $N=4$ it leads to a $(4,4)$
supersymmetric theory on $\Sigma$.  In this paper we will mostly
concentrate on the case of pure $N=2$ and $N=4$ YM theory.
Extension of these to $N=1$ and to $N=2$ theories with matter
are presently under consideration.

\def\S{\Sigma}
\def\e{\epsilon}
\def\ee{\epsilon}

\def\O{{\cal O}}

\newsec{Reduction}
We now consider this reduction in more detail.  Let us first
concentrate on pure  YM theory on a
four-dimensional manifold $M^4 $, which has a product structure
$M^4 =\S \times C$.
Let us choose the metric on this manifold to be block diagonal
$g=g_{\Sigma} \oplus g_{C}$, where $g_{\Sigma}$ ($g_{C}$)
is the metric
on $\S$ ($C$).
The YM connection can be decomposed into
two pieces
$A=A_{\Sigma}+A_{C}$, where $A_{\S}$ ($A_{C}$) is the component of
$A$ along
$\S $
($C$).
To discuss the reduction to 2d let us rescale the metric  $g_{C}
\rightarrow
\ee g_{C}$ on $C$.
Under this transformation different terms in the  action scale
differently
\eqn\act{\eqalign{S={1 \over 4 e^2}\int_{M^4 }{\rm Tr}\bigg[ {1 \over
\ee}F_{C} \wedge
\ast
F_{C} &+
 2(d_{C}A_{\Sigma}-D_{\Sigma} A_{C}) \wedge \ast
(d_{C}A_{\Sigma}-D_{\Sigma}
A_{C})
\cr &+  \ee  F_{\Sigma} \wedge \ast F_{\Sigma}\bigg]~.}}
Operation   $\ast$ is defined with respect to unrescaled metric
$g_{\Sigma} \oplus g_{C}$ and
 $D_{\S }=d_{\S }-i[A_{\S } ,\cdot]$.

In the limit $\ee \to 0$ the first term in the action enforces the
component
$A_{C}$ to be flat
($F_{C}=0$), while the second term gives rise
to the $\sigma$-model action. The last term produces the corrections
of
order
$O(\ee)$
that are irrelevant in the limit $\ee \to 0$.

We will denote the moduli space of flat
connections on $C$ by ${\cal M}(C)$.
In order to specify  the flat connection $A_{C}$
on $\S \times C$ one should specify
a map $X: \S \to {\cal M}(C)$. In this notation the flat
connection becomes
\eqn\con{A_{C}(w,\bar{w}, z, \bar z)=A_{C}(w,\bar{w} | X(z, \bar
z)),
}
where $z, \bar z$ ($w,\bar{w}$) are complex coordinates on $\S $
($C$).

The flatness condition $F_{C}=0$ implies that operator $D_{C}$ is
nilpotent,
$D_C^2=0$.
The tangent space to the moduli space of flat connection ${\cal
M}(C)$ is
given by $D_C$
cohomology $H^1 (C, {\cal G})$.
We will always choose representatives that
satisfy harmonicity condition $D_{C}^{\mu} \alpha_{\mu}=0$,
which is just the
gauge fixing condition.
The variation of the flat connection $\delta A_{C}$ can be
decomposed
with
respect to some basis
$\{ \alpha^I\} \subset  H^1 (C, {\cal G})$ modulo the gauge
transformation
\eqn\var{{\partial A_{C} \over \partial X^I}=\alpha_I+ D_{C} E_I~}
where $E$ defines the connection on the moduli space ${\cal M}(C)$
(similar
construction appears in \ref\hs{J.
Harvey and  A.
Strominger,
Comm. Math. Phys. {\bf 151} (1993) 221}, see also Appendix A).

The moduli space of flat connections
${\cal M}(C)$ is a K\"ahler manifold with the K\"ahler form and the
metric
given as follows
\eqn\kahl{\omega_{IJ}= {\rm Tr} \int_{C}\alpha_I\wedge
\alpha_J~~{\rm
and}~~~
G_{IJ}= {\rm Tr} \int_{C}\alpha_I \wedge \ast \alpha_J~.}
It is convenient to use the  complex coordinates
$X^i$ and
$X^{\bar{k}}$
on ${\cal M}$.

The action \act\ is essentially quadratic in $A_{\S}$, ignoring the
terms of
order $O(\e)$.
Moreover  the action does not depend on the derivatives of  $A_{\S}$
with
respect to the coordinates on $\S $. Hence $A_{\S}$ plays the role of
an
auxiliary field.
Therefore one can attempt to integrate out
 $A_{\S}$.
This
can be done if the connection on $C$ is irreducible, which would
allow
us to invert the Laplacian $D_{\overline w} D_w$ on $C$:
\eqn\Asigma{A_\S=E_i \partial_\S X^i +E_{\bar{k}}\partial_\S
X^{\bar{k}}~.}
If the gauge field on $C$ is reducible
the Laplacian has zero modes which would give rise to additional
degrees
of freedom on $\Sigma$ (and in
particular dropping the $O(\epsilon )$ terms in \act\ cannot be
justified
in such cases).
These additional degrees of freedom are described by residual gauge
theory on $\S$.
Moreover
if the dimension of the residual
gauge symmetry jumps as we
move on $\cal M$
the resulting 2d theory on $\Sigma$ would be very complicated.  This
happens
for example if we consider flat $SU(N)$
gauge fields on $C$.  However if we consider $SU(N)/Z_N$ gauge
theory
and restrict the path-integral to
the subsector where we turn
on a non-trivial `t Hooft magnetic flux on $C$, then the
connection on $C$ is irreducible for all ${\cal M}$, the gauge
group is completely broken and $A_{\S}$ can be integrated out.
We will mainly concentrate on this case, but comment about some
aspects
of the more general case below.

Substituting the flat connection $A_C$ and the expression for
$A_\S$ (eq. {\Asigma}) into
the action \act\ one gets the $\sigma$-model action of the standard
form
\eqn\Saction{S={1\over 2e^2}\int_\S  d^2 z  ~G_{i\bar{k}} (\partial_z
X^i
\bar{\partial}_{\bar{z}}
X^{\bar{k}} +
\bar{\partial} _{\bar{z}}  X^i \partial_z X^{\bar{k}}) ~.}
It is also easy to see that
turning on the $\theta $ angle for the YM is equivalent to turning
on a $B-$field in the direction of the K\"ahler class.
In this way we see that $\tau =i/4\pi e^2 + \theta /2\pi$ is now
playing
the role of the complexified K\"ahler modulus of this
$\sigma$-model\foot{
To fix the gauge at the quantum level we have to introduce the gauge
ghosts.
In the semiclassical approximation in the limit $\epsilon \to 0$ the
integration over quadratic fluctuations of the gauge field
(orthogonal to zero
modes) near the flat connection $A=A_C$  produces ${\rm det}^{-2}
\Delta$, where $\Delta$ is the covariant Laplacian on $C$, while
the integration over the ghost fields gives rise to the factor
${\rm det} \Delta$.
Combining together the determinants modifies the action
of two-dimensional $\sigma$-model.}.

The moduli space of
holomorphic instantons for this
$\sigma$-model
can be shown to coincide \ref\sal{S.~Dostoglou, D.~Salomon {\it
Instanton
homology and symplectic fixed points}, preprint 1992} with
the the moduli space of self-dual connections of the
4d YM
theory in the limit $\e  \to 0$.
In particular one can view anti-self-dual connections
as holomorphic connections (whose curvature vanish
in the $(2,0)$ and $(0,2)$ directions)
which satisfy $g^{i\bar j}F_{i\bar j}=0$.
The latter condition in the limit  $\e \to 0$ becomes
$F_C =0$, whereas the holomorphicity
of the connection is equivalent to holomorphic
instantons of the 2d theory.

Now consider the dimensional reduction of the topological
YM
theory which is the twisted version of the $N=2$ $d=4$
supersymmetric
Yang-Mills theory \ref\Wi{E.Witten, Comm. Math. Phys. {\bf 117},
(1988) 353}.  In this case one ends up with the
$(2,2)$ supersymmetric $\sigma$-model on ${\cal M}$.
It is convenient to formulate this model in the complex notation.
In the bosonic sector of $N=2$ SYM theory
there is a scalar field $\phi$ in addition to Yang-Mills connection
$A$.
The fermionic fields are the
following:  a
scalar $\eta$, a self-dual two form that can be decomposed to
a scalar $\lambda$ and
$(2,0)$ and $(0,2)$ forms $\lambda_{zw}$ and
$\lambda_{\bar{z}\bar{w}}$, and a $1$-form with the components
$\chi_z$,
$\chi_w$,  $\chi_{\bar{z}}$, $\chi_{\bar{w}}$.
Since the action is linear in fermionic fields
$\lambda$, $\eta$,
$\chi_z$ and $\chi_{\bar{z}}$, one can integrate them out.
 Such an integration  gives rise to the
following
constraints:
$D_w\chi_{\bar{w}}=0$, $D_{\bar{w}}\chi_w=0$,
$D_w\lambda_{\bar{w}\bar{z}}=0$,
$D_{\bar{w}}\lambda_{wz}=0$.
These fields are cotangent to the moduli space ${\cal
M}(C)$ of
flat connections on $C$.
Therefore in the basis $\alpha_{i\bar{w}}$, $\alpha_{\bar{k} w}$ they
can be
represented
as linear combinations
\eqn\ferm{\chi_{\bar{w}}=
\chi^i\alpha_{\bar{w}i},~\chi_w= \chi^{\bar{k}}
\alpha_{w{\bar{k}}},~\lambda_{\bar{w}\bar{z}}=
\rho_{\bar{z}}^i\alpha_{\bar{w}i},~\lambda_{wz}=\rho_z^{\bar{k}}
\alpha_{w{\bar{k}}},}
where $ \chi^i$, $\chi^{\bar{k}}$, $\rho_{\bar{z}}^i$ and
$\rho_z^{\bar{k}}$
are two dimensional fermionic fields on $\S $.
The action is also quadratic in scalar fields $\phi$ and
$\bar{\phi}$
and does
not depend on the derivatives of these fields with respect to
coordinates on $\S $ (in the leading order $\e \to 0$).
Notice that self-interaction $\phi^4$
is suppressed in the limit $\epsilon\to 0$.
Therefore one can just solve the equations of motion
for $\phi$ and $\bar \phi$
\eqn\exphi{\phi=\chi^{\bar{k}} \chi^i \Phi_{\bar{k}i},~~~
\bar{\phi}=g^{z\bar{z}} \rho^{\bar{k}}_z \rho_{\bar{z}}^i
\Phi_{\bar{k}i}  ,}
where $\Phi_{\bar{k}i}$ is the curvature on the principal bundle
on ${\cal M}$ (see Appendix A).

Similar to the above non-supersymmetric model we integrate over
components
$A_\S$ of the gauge connection.
Semiclassically $A_{\S}$ is given by an expression \Asigma\
plus a
bilinear combination of the fermionic fields.
Thus the integration over the field $\phi$ and $A_{\S}$ results in a
four-fermionic interaction in the Lagrangian.

At this stage, due to the underlying $(2,2)$ supersymmetry
and having identified the target space one can directly
write the $\sigma$-model action for the reduced theory. However
it is instructive to check that it really results from this
reduction.
In particular in
the quantum theory one has to introduce the gauge ghosts and
integrate
over
the quadratic fluctuations near the solutions of the classical
equations of
motion.
It is easy to check that the determinants of the Laplace operators
cancel due to
supersymmetry in
contrast to the above non-supersymmetric case.
Therefore, as expected, we get
a supersymmetric twisted
$\sigma$-model on ${\cal M}$  ({\bf A} model) with the
standard action\foot{The {\bf A} twisting is
inherited from four dimensions.
If we consider the partial twisting of the four
dimensional theory described above, we  would obtain
the untwisted $\sigma$-model on ${\cal M}$.}
 \ref\Amodel{E.Witten, in {\it Essays on Mirror Manifolds}, ed by S.
T. Yau,
International Press, 1992}
\eqn\Amodel{\eqalign{S= {1\over e^2}\int_\S {\rm d}^2z~
\bigg[G_{i\bar{k}}({1\over
2}\partial_z X^i
\bar{\partial} _{\bar{z}} X^{\bar{k}} +{1\over 2}
\bar{\partial} _{\bar{z}} X^i \partial_z X^{\bar{k}}&+
i\rho^{\bar{k}}_z
\overline{D}_{\bar{z}}\chi^i+
i\rho^i_{\bar{z}}D_z\chi^{\bar{k}}) \cr  &- R_{i\bar{k} j\bar{l}}
\rho^i_{\bar{z}} \rho^{\bar{k}}_z \chi^j \chi^{\bar{l}}\bigg],}}
 where $D_z\chi^i =\partial_z\chi^i +\chi^j \Gamma^i_{jk}\partial_z
X^k$,
 $\overline{D}_{\bar{z}}\chi^{\bar{i}}
=\bar{\partial}_{\bar{z}}\chi^{\bar{i}}
+\chi^{\bar{j}}\Gamma^{\bar{i}}_{\bar{j}\bar{k}}\bar{\partial}_{\bar{z
}}
X^{\bar{k}}$.

The anomaly in the fermion number is the same for the
original 4d
topological theory and for the $\sigma$-model.
In the case of $SU(N)$, in particular the $c_1({\cal M})=N h_2$
(where $h_2 \in H^2({\cal M},{\bf Z})$), in accord with the $U(1)$
`ghost'
number violation for
the $N=2$ $SU(N)$ theory.

The physical operators of this $\sigma$-model are given by the BRST
cohomology.
Let us consider the relation between the physical operators in
$\sigma$-model
and
those of the 4d topological YM theory.
It is easy to check that under dimensional reduction
the local physical operator of ghost number 4 of the 4d
topological theory
$\O={\rm Tr} \phi^2$ becomes
\eqn\hfour{\O\to  b=\chi^i \chi^{\bar{k}} \chi^j\chi^{\bar{l}}
R_{i\bar{k} j\bar{l}}\in H^4({\cal M}) .}
Actually ${\rm Tr}\phi^2 \to b$ is true only classically.
As we will see later, at the quantum level they differ by a
$c$-number.  Its descendant, a non-local physical operator $\int_C
\O^{(2)}
={\rm Tr}\int_C d^2 w~(F_{w\bar{w}}  \phi + \chi_w\chi_{\bar{w}})$
with the ghost number 2 becomes a local operator which represents
a K\"ahler class
\eqn\htwo{I(C)=\int_C \O^{(2)} \to a=\omega_{i\bar{k}}\chi^i
\chi^{\bar{k}}\in
H^2.}
The fermionic local operators $\gamma^i \in H^3$
correspond to the first descendant of the operator $\O$
integrated over 1-cycles $c_i$ on $C$:
\eqn\aibi{{\gamma_i}= \int_{c_i} (\chi^i\chi^{\bar{k}}
\chi^{\bar{l}}~
{\rm Tr}\Phi_{i\bar{k}} \alpha_{w \bar{l}} dw+
\chi^i \chi^{\bar{k}} \chi^j ~{\rm Tr}\Phi_{i\bar{k}} \alpha_{\bar{w}
j}
d\bar{w}).}
All the rest of the physical operators in the 4D topological
YM theory
become the descendants of these operators.\foot{Of course in the
topological
$\sigma$-model there are more physical operators. However the above
listed operators generate all the rest.}

Consider now the dimensional reduction of the $N=4$
SYM theory.
It is convenient to consider the {\it partially}-twisted version of
this
theory.
In the complex notation the bosonic content of the model is the
following:
the gauge field $A$, two complex scalar bosons $\phi$ ($\bar{\phi}$)
and $\varphi$
($\bar{\varphi}$) and $(1,0)$ and $(0,1)$ forms on $C$ denoted by
$\phi_{w}$
and
$\phi_{\bar{w}}$
respectively\foot{The twist that we use is the
partial twisting corresponding to that
used in
ref.\ref\Sdual{C.Vafa and E.Witten, A Strong Coupling Test of
S-Duality.
Preprint HUTP-94/A017, IASSNS-HEP-94-54. hep-th/9408074}.}.
These non-scalar bosonic fields appear because  twisting is
performed with
conserved  current that includes a bosonic contribution.
The fermionic fields are doublets with respect to the $SU(2)$ global
group
which is the unbroken subgroup of $SU(4)$ corresponding to $N=4$
supersymmetry
(the BRST
charges are doublets with respect to this global group).
There are the following fermionic fields: two scalars (on $C$)
$\eta_-^a$ and
$\lambda^a_{-}$,  $(1,0)$ and $(0,1)$ forms $\lambda^a_{w-}$ and
$\bar{\lambda}^a_{\bar{w}-}$,  two vectors represented
(after contracting with metric) by
$\chi^a_{w+}$,
$\bar{\chi}^a_{\bar{w}+}$, and additional scalars on $C$ denoted
by $\chi^a_+$ and
$\bar{\chi}^a_{+}$.
Here the indices $a=1,2$ correspond to the doublet representation of
the unbroken
$SU(2)$ global group,
the vector indices $w$ and $\bar{w}$ correspond to the surface $C$,
and the
indices $\pm$ stand for (right-) left-handed spinors indices on
$\S $.

The dimensional reduction here is slightly different from that of
above cases
in the
following respects.
First,  some of the bosonic fields which are scalar fields in the
untwisted
theory become 1-forms in the twisted model. Therefore their kinetic
term
is not
suppressed as $\epsilon\to 0$ and may still correspond to
propagating degrees of freedom in the dimensionally reduced theory.
Second, there are unsuppressed terms in the Lagrangian which
describe
$\phi^4$
interactions of the bosonic fields.
In the limit $\e \to 0$ the equations of motion reduce to
\eqn\hitreduce{F_{w\bar{w}}= -i[\phi_{\bar{w}},\phi_{w}],~~~D_w
\phi_{\bar{w}}=0, ~~~D_{\bar{w}} \phi_{w}=0
.}
This set of equations coincides with the Hitchin's equations for
`stable pairs'
\ref\hit{N.J.Hitchin,
The self-duality equations on a Riemann surface, {\it Proc. Lond.
Math. Soc.}
{\bf 55} (1987) 59.}.  The moduli space ${\cal M}^H$ of solutions
to Hitchin equations is the
target space of supersymmetric 2d $\sigma$-model.
Roughly speaking, one may think of Hitchin's space as partial
compactification of the cotangent bundle to the moduli space of
flat connections.
As expected, ${\cal M}^H$ turns out to be a hyperK\"ahler  manifold
of dimension $dim_{\bf C}{\cal M}^H=6g-6$ \hit.
This implies that $\sigma$-model
has $(4,4)$ superconformal symmetry.
These facts are discussed in appendix B\foot{There are interesting
generalizations of Hitchin equation when we consider other theories.
For example for $N=1$ theories with
$SU(N)$ gauge group and with $2N$ flavors a similar equation
appears where there are $N$ holomorphic $\phi$'s in the above
equation
appearing in the fundamental representation of $SU(N)$.
This is a generalization
of Vortex equations
studied by mathematicians \ref\vort{S. B. Bradlow, J. Diff. Geom {\bf
33}
(1991) 169-213\semi
S.B. Bradlow, G.D. Daskalopoulos, O. Garcia-Prada and R. Wentworth,
Stable Augmented Bundles over Riemann Surfaces, Symposium on Vector
Bundles in Algebraic Geometry, Durham 1993, London Math. Soc. Lecture
Notes Series, CUP (in press).
}.  These generalizations
of Hitchin space are currently under investigation.}.

\newsec{Applications}

\subsec{Aspects of Target Spaces ${\cal M}$ and ${\cal M}^H$}

In this section we make some comments about ${\cal M}$ and ${\cal
M}^H$.
For concreteness let us concentrate on $SU(2)/Z_2=SO(3)$.
We denote  by $g$ the genus of $\S$ and by $h$ the genus of $C$.
The SO(3) bundles on $\Sigma\times C$  are characterised  by the
instanton
number (the first Pontryagin class) and the
`t Hooft magnetic flux (the second
Stiefel
-Whitney class) $z\in H^2(\Sigma\times C,{\bf Z}_2)$.  The magnetic
flux has
$4gh+2$ components corresponding to decomposition  $H^2(\Sigma\times
C,{\bf
Z}_2)=H^2(\Sigma,{\bf Z}_2)+H^1(\Sigma,{\bf Z}_2)\otimes H^1( C,{\bf
Z}_2)+H^2(
C,{\bf Z}_2)$. To avoid complications as discussed before
 in this paper we mainly consider
gauge field configurations
with nonzero magnetic flux  $z(C)=1$ through the ``small''
surface
$C$. Then ${\cal M}$ can be essentially identified with
the space of representations of the group
$\tilde{\pi}_1(C)=\langle
a_i,\,b_i\, |\prod_{i=1}^{h}a_ib_ia_i^{-1}b_i^{-1}=-1\rangle$
in $SU(2)$. (An
unusual $-1$
in the right hand side appears exactly because of  the nonzero flux
through
$C$.) This space is smooth and compact.
Note that flipping
the signs of $a_i$ and $b_i$ does not change the element they
correspond to on $SO(3)$.
Thus to be more precise the space that appears for the target space
in the
$SO(3)$ theory is an {\it orbifold} of this space.
We consider on  ${\cal M}$ a group $G=H^1( C,{\bf Z}_2)$
which acts by flipping the
signs of
$a_i$ and $b_i$.
This action is not free and each element  $\alpha \in
G$ fixes an
$(h-1)$-dimensional complex torus $T_\alpha^{2(h-1)}$. The
$\sigma$-model
which corresponds to 4d $N=2$ SYM has as a target space
${\cal M}/G$. The path-integral sectors of the orbifold
$\sigma$-model on the
worldsheet
$\Sigma_g$ are classified by the boundary conditions on
$A_i$ and
$B_i$ cycles of $\Sigma_g$, i.e. by the elements of
$H^1(\S,G)\approx
H^1(\Sigma,{\bf Z}_2)\otimes H^1( C,{\bf Z}_2)$.  As one can  see in
the  4d
language by using the definition of `t Hooft magnetic flux
the  boundary conditions are equivalent to the choice of
$4gh$
components of the magnetic flux  $z$ in  $H^1(\Sigma,{\bf
Z}_2)\otimes H^1(
C,{\bf Z}_2)$.  There remains a flux  $z(\Sigma )$ through the
worldsheet
which is either zero or one.
Apriori since we have not fixed it, we expect that the $\sigma$-model
sums over both allowed values.  The instanton number
$p$ of the SYM theory is $p=-z^2/4$ mod 1 differs by a factor of $2$
from the
$\sigma$-model instanton number $k=2p=-z(\S)+ const$, where $const$
depends on the orbifold subsector. Therefore
 turning on and off the $z(\Sigma)$
for each orbifold subsector
shifts the parity of the instanton number.  We can thus isolate
the contributions corresponding to $z(\Sigma )$ on and off
in the 2d $\sigma$ model.

The same story repeats for the Hitchin space ${\cal M}^H$.
The $\sigma$-model appearing in the physical theory
is an orbifold of this space corresponding to the group $G$ described
above.  In fact ${\cal M}^H$ may be viewed \hit\ as a certain
partial compactification of $T^* {\cal M}$ and the relevant sigma
model  is the quotient of this space by $G$.

\subsec{N=2 Application}
We now discuss the implications of the above reduction
for the $N=2$ YM theory.  In the $N=2$ case in the fully
twisted version
we can use the computation of Donaldson invariants in the 4d
theory to compute quantum cohomology ring on ${\cal M}$\foot{
If we consider the case $C=S^2$, the gauge fields on
$C$ would be frozen and thus we obtain an effective theory
on $\Sigma$ which is just the $N=2$ YM.  In this case the Donaldson
observable of 4d get mapped to Donaldson observables
of the 2d YM which has been shown
\ref\WiYM{E.~Witten,
{\it Gauge theories revisited}, J.~Geom.~Phys.~{\bf 9} (1992) 303}\
to compute {\it classical} cohomology ring of ${\cal M}(\Sigma )$.
If, on the other hand,
 we take $S^2$ to be large and $\Sigma$ to be small, as we
have argued, we obtain the {\it quantum} cohomology ring of
${\cal M}(\Sigma )$.  This is not a contradiction, as it is
known that Donaldson theory is anomalous when $b_2^+=1$
and the topological amplitudes depend on the choice
of metric, as is
the case here.  Similarly, in the context of $N=4$ the topological
reduction on
$S^2$ gives $N=4$ YM on $\Sigma $ which
should thus enjoy $S$-duality.}.

The classical cohomology of  ${\cal M}$ is well studied both by
mathematical
\ref\AB{M.~Atiyah and  R.~Bott {\it The Yang Mills Equations over
Riemann
Surfaces}, Phil. Trans. R. Soc. Lon {\bf A 308}, 523-615
(1982)}\ref\Ki{F.~Kirwan, J. Amer. Math. Soc. {\bf 5} 853-906
(1992)}\ref\Th{M.~Thaddeus, Invent Math. 117, 317-353 (1994)}\ and
physical
\WiYM\
methods.
 The ring  $H^*({\cal M})$ is generated
by the
elements  $a$, $\{\gamma_i \}_{i=1}^{2h}$, $b$ of degrees 2, 3 and 4
respectively. In our language these generators appear, as discussed
before,
as $a=\int
_{C}[{\rm Tr}\,
\phi^2]^{(2)}$,
$\gamma _i=\int _{c_i}[{\rm Tr}\, \phi^2]^{(1)}$ and
$b={\rm
Tr}\, \phi^2$.
Note that although {\it nonlocal} in 4d SYM,  the operators $a$ and
$\gamma_i$
become {\it local} after the reduction to 2d $\sigma$-model.  The
modular group of
$C$ acts on $\gamma_i$.
The intersection numbers (correlation functions) are
modular-invariant,  hence
they can be
computed in
terms of  the modular-invariant subring  $H^*({\cal M})^{inv}$,
generated by
$a,$ $b$ and $c=\sum J_{ij}\gamma_i \gamma _j$ where $J_{ij}=c_i\cap
c_j$ is
the intersection form on $C$.
There are three relations $R_1^h=0,R_2^h=0,R_3^h=0$ in $H^*({\cal
M})^{inv}$
in degrees $2h$,  $2h+2$, $2h+4$ respectively. And the relations for
$h+1$ can
be expressed in terms of ones for $h$ by the recursion relations of
\ref\TS{B. Seibert and G. Tian,
{\it Recursive relations of the cohomology ring of the moduli spaces
of stable
bundles},
Preprint, October 21, 1994}, \ref\NS{A.~King and P.~Newstead, {\it On
the
Cohomology
ring of the Moduli Space of rank 2 vector bundles on a curve},
preprint 1994}:
$R_1^{h+1}=aR^h_1+h^2R^h_2$,
$R_2^{h+1}=bR^h_1+{2h\over h+1}R^h_3$,  $R_3^{h+1}=cR^h_1$. (Formally
$R_1^1=a,R_2^1=b,R_3^1=c$. )

The chiral ring  of the topological $\sigma$-model on ${\cal M}$ is a
quantum
deformation of  $H^*({\cal M})$.  It suffices to find the
deformation of
$H^*({\cal M})^{inv}$ because it gives the full information about
the modular
invariant correlation functions of the $\sigma$-model. This
deformation
is
generated by $a$, $b$, $c$ and $q={\rm exp} (2\pi i \tau)$, where $q$
counts the instantons and has a formal degree 4 (because
$2c_1({\cal M})=4$).  There are three relations
$Q_1^h=0$,
$Q_2^h=0$ and $Q_3^h=0$ which are reduced to $R_1^h=0$, $R_2^h=0$ and
$R_3^h=0$ for
$q=0$.  We will be able to find the quantum relations using the
results of 4d
Donaldson theory. But first we need to describe  more precisely the
correspondence between 4d  $SO(3)$ SYM and 2d $\sigma$-model.

The group $G$ acts trivially on $H^*({\cal M})$.  Thus the chiral
ring of the
{\it untwisted} sector of  orbifold ${\cal M}/G$ is the same as
the chiral
ring  of  the $\sigma$-model on ${\cal M}$.  It allows us to draw
conclusions
about the quantum ring of that $\sigma$-model from the correlation
function
$\sum_z\langle e^{\alpha a +\beta b}\rangle_z$ computed in $SO(3)$
Donaldson
theory on $\Sigma \times C$. Two  magnetic fluxes corresponding to
untwisted
boundary conditions have zero components in $H^1(\Sigma,{\bf
Z}_2)\otimes H^1(
C,{\bf Z}_2)$ and differ by the value of  $z(\Sigma )$.  When $\Sigma
=T^2$ is
a complex 1-torus,
\eqn\donsig{\sum_{z(\S)}\langle e^{\alpha a +\beta b}\rangle^{T^2
\times
C}_z={\rm Str}_{\cal H} e^{\alpha a +\beta b}}
computes the (weighted
with
signs) sum over the spectrum of $a$ and $b$.  In principle, because
of the
signs,  the contribution of some eigenvalues could be cancelled  off
completely. We make a minimal assumption that  it does not  happen
and that we
can read the whole spectrum off   ${\rm Str}_{\cal H} e^{\alpha a
+\beta b}$.
The correlation function $\langle e^{\alpha a +\beta b}\rangle^{T^2
\times
C}_z$ can be obtained using the results of \ref\WiI{E.~Witten  {\it
Supersymmetric Yang Mills theory on a four-manifold}, preprint
IASSNS-HEP-94/5,
1994.}
or
\ref\WiII{E.~Witten {\it Monopoles and four-manifolds}, preprint
IASSNS-HEP-94-96}\foot{In fixing the overall normalization
we have to be a bit careful:  Taking into account that we are
discussing
the $SO(3)$ as opposed to $SU(2)$ case modifies the overall
coefficient
of \WiI \WiII\ by a factor of $2^{-b_1}$ where $b_1= 2(g+h)$.  In
addition
in the $\sigma$-model at genus $g$ we have the orbifold symmetry
factor
$1/({\rm dim} G)^g=2^{-2gh}$;  So we have to multiply the overall
normalization of \WiI \WiII\ by $2^{-2g-2h+2gh}$, which up to a
redefinition of the string coupling constant on
$\S$ leaves us with a factor
of $2^{-1}$.}:
\eqn\tor{\eqalign{
\langle e^{\alpha a +\beta b}\rangle^{T^2 \times C}_z=(-1)^{z(\Sigma
)(h-1)}\bigl[&e^{-\lambda \beta}(e^{-\alpha x}-(-1)^{z(\Sigma
)}e^{\alpha
x})^{2h-2} +\cr
&i^{-z^2}e^{\lambda \beta}(e^{-i\alpha x}-(-1)^{z(\Sigma
)}e^{i\alpha
x})^{2h-2}\bigr]/2
}}
where $\lambda =8q$ and $x=2i{\sqrt q}$
(these normalizations are
consistent
with the one used by Donaldson \ref\Don{S. K. Donaldson, {\it Floer
homology
and algebraic geometry}, preprint 1994}).
Summing \tor\ over two values of  $z(\Sigma )$ one sees that $b$ has
two
eigenvalues $\pm \lambda $ and $a$ has $2g-1$
eigenvalues $\pm (0,\,
 2ix, \,  4x,\,  12ix,\,  16x,\,\ldots)$. As for the third
generator
$c$,  since it is bilinear in fermions it is nilpotent ($c^{g+1}=0$)
so the
only eigenvalue is $0$.
The
spectrum together with the condition that the quantum ring relations
are
$q$-deformations of the classical ones determine the relations
$Q_1^h$,
$Q_2^h$,  $Q_3^h$ completely. We put formally $Q^1_1=a, Q^1_2=b-8q,
Q^1_3=c$.
A straightforward analysis shows that
the recursion relations are modified in a simple way:
\eqn\qreodd{
Q_1^{h+1}=aQ^h_1+h^2Q^h_2,\ \
Q_2^{h+1}=(b+(-1)^{h-1}8q)Q^h_1+{2h\over h+1}Q^h_3,\ \
Q_3^{h+1}=cQ^h_1}
In particular, for genus $2$ the relations are
$Q^2_1=a^2+b-8q,~~Q^2_2=ba+8qa+c,~~Q^2_3=ca$.
Up to a redefinition of generators $a \to h_2,\,b \to -4h_4+4q,\, c
\to
4(h_6-qh_2)$
 these relations coincide with those
found by Donaldson (for $g=2$) \Don, where $h_2,\,h_4,\, h_6$ are
generators
of {\it integral} cohomology.
Already this example shows that at the quantum level the
$\sigma$-model
operator  $b={\rm Tr}\phi^2$ is shifted by $\propto q\cdot 1$ and the
operator
$c$ is shifted by $\propto q\cdot a$ from the rational cohomology
generators
they used to be classically.  As discussed before
this possibility is allowed as in going from classical
to quantum identification we had to consider  composite operators
and the definition of composite operator may receive instanton
correction.

The simplicity of \qreodd\ should not be misleading. The very
existence of
quantum deformation
consistent with the Donaldson spectrum is by no means obvious.
Moreover,
one can find the full quantum cohomology ring starting from the
modular
invariant
subring defined by \qreodd.  Taking into account the fermionic
contributions
(odd cohomology)
one can compute the partition function on a torus (${\rm Str}_{\cal
H} e^{\alpha a +\beta b}$).
We explicitly checked the relation \donsig\ for  small genus $g=2,3$.
For example for $g=3$ the one loop partition function
\eqn\bos{\eqalign{{\rm Str}_{\cal H} e^{\alpha a +\beta b}=&
\Big( 2e^{\lambda \beta}(e^{2 \alpha i x}+ e^{-2 \alpha i x})
+e^{-\lambda \beta} (18+e^{4\alpha x}+ e^{-4\alpha x})\Big) -  \cr
&~~~~~~ \Big( 6 e^{\lambda \beta} (e^{2 i \alpha x}+ e^{-2 i \alpha
x}) +12
e^{-\lambda \beta}\Big) \cr}}
These two brackets are bosonic and fermionic contributions
respectively.
$14$ out
of the $18$
bosonic states corresponding to eigenvalues $(b=-\lambda, a=0)$
correspond to the bilinears in fermions.

For the  worldsheets $\Sigma_g$ of genus $g>1$ the 4d SYM   \WiII\
gives the
answer
\eqn\hig{\eqalign{
\langle e^{\alpha a +\beta b}\rangle^{\Sigma_g \times
C}_z=&(-1)^{z(\Sigma
)(h-1)}\bigl[e^{-\lambda \beta}(e^{-\alpha
x(h-1)}+(-1)^{(g-1)(h-1)}e^{\alpha x(h-1)}) \cr
&+i^{-z^2}e^{\lambda \beta}(e^{i\alpha
x(h-1)}+(-1)^{(g-1)(h-1)}e^{i\alpha
x(h-1)})\bigr]/2
}}
One may find the derivation of this formula in Appendix C.
Taking the sum over $z(\Sigma_g)=0,1$  one sees that only two out of
$2h-1$
eigenvalues of $a$ contribute, those with the maximal absolute value
$4(h-1)\sqrt{q}$. This is consistent with the fact that only
for these values of $a$ the above ring relations give non-degenerate
eigenvalues of operators $(a,b,c)$.  The higher
genus amplitude is obtained by inserting the handle operator $H$
raised to the power of $g-1$, and the handle operator is essentially
the mass operator which vanishes at degenerate
points as there is no mass gap.  If $g>1$ at the degenerate
eigenvalues
$H^{g-1}$ vanishes leaving us with the contribution of
the two non-degenerate eigenvalues.

Having described the untwisted sector of  our orbifold
$\sigma$-model,
we can
turn to the twisted sectors.  The Hilbert space ${\cal H}_\alpha$ of
the
$\alpha$-twisted sector, $\alpha \in G$, consists of  $G$-invariant
part  of
the cohomology of the complex $(h-1)$-dimensional torus
$T^{2(h-1)}_\alpha$
fixed by $\alpha$. The fermionic numbers of all the vectors from
${\cal
H}_\alpha$ should be shifted up by $2(h-1)$  to match those from the
untwisted
sector. The group $G$ acts on $T^{2(h-1)}_\alpha$ by translations and
reflections, so the $G$-invariant part of cohomology coincide with
the even
part.  Finally, ${\cal
H}_\alpha=H_{even}^{*+2(h-1)}(T^{2(h-1)}_\alpha)$ and
${\rm dim}\,{\cal H}_\alpha=2^{2h-3}$.
Note that since ${\cal H}_\alpha$ is bosonic  its dimension is
computed by the
Witten index ${\rm Str}_{{\cal H}_\alpha}1$. One can
check that
the 4d computation gives just the right answer  $2^{2h-3}$ (the
computation
involves summation over $2^{2h}$ fluxes $z$ necessary to project onto
$G$-invariant states).  In fact we can do better, namely we can
examine the contribution of each twisted path-integral sector.
 The
orbifold partition function on torus with the boundary conditions
along $A,\,B$
cycles  twisted by ${\cal A} ,{\cal B} \in G$ computes, in absolute
value, the number
$F_{{\cal A} ,{\cal B}}$ of points fixed by both  $\cal A $ and
$\cal B$.  The formula
\tor\
leads to $F_{{\cal A} ,{\cal B}}=0$ if ${\cal A} \cap {\cal B} =0$
and
$F_{{\cal A} ,{\cal B}}=2^{2h-2}$ if  ${\cal A} \cap {\cal B}=1$ (we
use here
the cap
product $ \cap
$  in $G=H^1(C,{\bf Z}_2)$ and the identification $H^2(C,{\bf
Z}_2)={\bf
Z}_2$). This means that although the tori  $T^{2(h-1)}_{\cal A}$ and
$T^{2(h-1)}_{\cal B}$ have very small dimension --- $2{\rm dim}\,
T^{2(h-1)}_{\cal A}=4(h-1)<6(h-1)={\rm dim}\, {\cal M}$ --- they
intersect
one another!  This surprising conclusion is true and can be
explicitly checked
using the
definition of  ${\cal M}$ as a  space of representations of
$\tilde{\pi}_1(C)$.  In particular if ${\cal A} \cap {\cal B} =0$,
$\cal B$ acts as translation on $T^{2h-2}_{\cal A}$ and if
${\cal A} \cap {\cal B} =1$ it acts as a total reflection giving
$2^{2h-2}$ fixed points.
\subsec{N=4 Application}

Now we turn to the discussion of aspects of the reduced
$N=4$ YM in two dimensions.
As we discussed before the two dimensional theory we have obtained
is a supersymmetric $\sigma$-model on the Hitchin space ${\cal M}^H$
which
is a hyperK\"ahler manifold.
Since ${\cal M}^H$ is a smooth hyperK\"ahler manifold the
corresponding sigma
model
is a superconformal theory\foot{Just as in the $N=2$ case, for the
$SO(3)$ theory the actual target for the
$\sigma$-model
is the quotient ${\cal M}^H/G$ which is also a nice superconformal
theory
if ${\cal M}^H$ is.  Moreover one can easily extend
the $S$-duality discussed in this section in the context of ${\cal
M}^H$
to that for ${\cal M}^H/G$ which effectively has the matrix
$D$ discussed in this section replaced by $D^{-1}$. Aspects of the
sharpened
version of the $S$-duality conjecture can be verified in this
context.}, which
is in accord with the fact that one
expects the four dimensional theory to be superconformal as well.

 Since the coupling constant $\tau$ of the 4d YM theory
gets identified with the unique complexified K\"ahler class for
${\cal M}^H$,
the Montonen-Olive conjecture \ref\om{C.Montonen and D Olive, Phys.
Lett,
B72 (1977) 117 }\
for the 4d N=4 YM,
 gets translated to the
modularity properties of the topological $\sigma$-model with respect
to the
K\"ahler moduli $\tau$. In particular for $SU(N)$, the moduli space
for
$\tau$ should be a fundamental domain for the subgroup $\Gamma_0(N)$
(with
lower
off-diagonal entry being 0 mod $N$)
of $SL(2,{\bf Z})$ (see \Sdual\ and note that $\Sigma \times C$ has
even quadratic form on $H^2$).
The $S$-duality conjecture in 4d thus gets translated to a
$T$-duality
for this
2d $\sigma$-model.  However
for $\sigma$-models we basically understand
how $T$-duality may arise and thus we may be able to shed some light
on the
S-duality
in 4d theories.  We will show why the Hitchin's $\sigma$-model has
$T$-duality.
Before doing this let us see why this map of
$S$-duality to $T$-duality is a resonable thing to expect.

In fact this is a natural
generalization of the $S$-duality for the abelian $N=4$ theory:
If we consider $SU(2)$ gauge group and choose
the internal space $C=T^2$, with a magnetic
flux turned on, the $\sigma$-model becomes trivial
(i.e. the Hitchin space is just a point).  However if
we don't turn on the flux, as discussed before we do not get a simple
reduction to a $2d$ theory as different $4d$
field configurations lead to
different regimes of the
reduced theory which are connected to each other in a
complicated way.
In one field regime which corresponds to large expectation values for
$\phi$,
i.e. the Higgs phase,
the theory reduces to a $U(1)$ gauge theory plus a
$\sigma$-model on the corresponding Hitchin space which in
this case is just the cotangent of the
moduli of flat connection (i.e. the cotangent of
the torus which characterizes the
holonomy
of the unbroken $U(1)$ along the $T^2$ modulo the Weyl action).  In
other words, as noted in \ref\por{L.~Girardello, A.~Giveon,
M.~Poratti, A.~Zafaroni, Phys.~Lett.~334B (1994) 331; and to appear.
}, the piece of the partition function
compactified on $T^2$, which grows like the volume of $\phi$,
can be easily extracted from this complicated effective theory and is
manifestly $S$-dual since for large $\phi$ the $S$-duality for the
non-abelian
theory gets mapped
to $S$-duality for the abelian theory.  In this context the field
configurations which wrap around the
$\sigma$-model torus get mapped to 4d field configurations
where there is a magnetic flux for this unbroken $U(1)$ and the
momentum modes are the dual configurations which are identified
with the electrical flux of the unbroken $U(1)$.  Thus the
$S$-duality
of the abelian theory gets mapped to $T$-duality\foot{The fact
that in this context the $S$-duality is equivalent
to the $T$-duality of toroidal compactification of the
reduced theory has been
independently noted in a recent paper \ref\mhs{J. Harvey, G. Moore
and
A. Strominger, {\it Reducing $S$-Duality to $T$-Duality},
hep-th/9501022}.}.
Note, however,
that it would be incorrect to ignore the other field configurations
which make contributions to the path-integral which
do not grow like the volume of $\phi$.  In fact it is relatively
easy to see that ignoring those would lead to a Witten index
for the $\sigma$-model which does not agree with that for
the 4d theory (which for $SU(2)$ is
1 for the $\sigma$-model and 10 \Sdual\ for the 4d theory).  Thus to
make
a really non-abelian test of $S$-duality we turn to the case
where genus of $C$ is greater than 1 and with `t Hooft magnetic
flux turned on.

   There is a description of ${\cal M}^H$
which is most
suitable for us \hit:  For any gauge group $G$,  ${\cal M}^H$ is a
fiber space
over
the complex space ${\bf C}^d$  where $d={\rm dim G}(h-1)$, whose
fiber is a
complex torus with
complex dimension $d$.  The complex structure of the torus varies
holomorphically
as we move in ${\bf C}^d$, but the K\"ahler structure of the torus is
fixed and
can
be identified with the K\"ahler structure of ${\cal M}^H$.  As we
move the base
point
we reach points where the fiber is a singular torus but the total
space is
still smooth.
The situation is a  generalization of the cosmic string solution
constructed
in \ref\gsvy{B. R. Greene, A. Shapere, C. Vafa and S.-T. Yau,
Nucl. Phys. {\bf B337} (1990) 1.}\ where the base was ${\bf C}^1$ and
the fiber
a complex one
dimensional
torus.  The basic strategy there was to use adiabatic approximation,
by
viewing the complex moduli as massless fields in ${\bf C}^1$ and to
construct a hyperK\"ahler
metric by adiabatically
varying the complex structure but with a fixed K\"ahler
structure of the torus.   Since the K\"ahler moduli is fixed for each
fiber,
this means that
the modular properties of the K\"ahler  moduli we will obtain,
as long as we can trust the adiabatic approximation, will still be
the same as
that for each fiber (as the
massless fields corresponding to varying it are turned off).
 The adiabatic approximation breaks down in the regions
where the fiber becomes singular--however as was the case in \gsvy\
and as is the case for Hitchin space the total
space is still a smooth hyperK\"ahler space and we thus obtain
an exact $(4,4)$ superconformal theory.
Even though we may not have trusted adiabatic approximation
for obtaining exact solutions, we do trust it as far as symmetries
are concerned.  Thus
the K\"ahler moduli $\tau$ which can be identified with that
of a non-singular fiber still enjoys the same modular properties
as that of each fiber.
Thus to find the modular properties of the K\"ahler parameter $\tau$
for ${\cal M}^H$ we have to
study
the modular properties of the K\"ahler modulus of the fiber torus.

Let us briefly explain why ${\cal M}^H$ has this toroidal
fiber structure.  For simplicity let us concentrate on $G=SU(2)$.
Let $b_{ww}={\rm det}\phi_w=-{1\over 2}{\rm Tr}\phi_w^2$. Then by
Hitchin equations \hitreduce, $\overline \partial b_{ww}=0$
whose solution can be identified with ${\bf C}^{3h-3}$, i.e. the
complex $3h-3$ dimensional space. Generically a point
of ${\bf C}^{3h-3}$ will correspond to a $b_{ww}$ with isolated
zeroes.
Let us concentrate on such a solution.  Away from the zeroes
of $b_{ww}$, $\phi_w$ determines a $U(1)$ subspace of $SU(2)$, by
the condition that $\Lambda=\phi_w/{\sqrt b_{ww}}=\pm 1$--more
precisely
we obtain a line bundle on the double cover $\hat C$ of $C$,
which has genus $4h-3$, branched over
the zeroes of $b_{ww}$.  Away from the branch points the gauge field
restricted to this $U(1)$ part is flat as follows from the fact that
${\rm Tr}F(\Lambda \pm 1)=0$ because ${\rm Tr}F=0$
and ${\rm Tr}F \phi_w= {\rm Tr}[\phi_w ,\phi_{\overline w}]\phi_w
=0$.
This line bundle will have delta function singularities at the branch
points that can be gotten rid of by tensoring with a fixed line
bundle
with opposite singularity.  The possible solutions to the Hitchin
equation will thus give rise to flat bundles on $\hat C$
which are parametrized by the Jacobian of $\hat C$, which can be
viewed as the allowed holonomies of the $U(1)$ gauge group
through the cycles of $\hat C$. However the allowed fluxes
are parametrized by the Prym subspace of the Jacobian, which
is the $3h-3$ dimensional complex torus which is odd under the $Z_2$
involution. This is because the involution on $\hat C$ exchanges
the line bundle with its dual.  We have thus given the description
of ${\cal M}^H$ as a toroidal fiber space over $C^{3h-3}$.  The
generalization
to $SU(N)$ is straightforward, with the base space being replaced by
the space of allowed holomorphic differentials $Tr\phi_w^j$,
where $j=2,...,N$,
and by the fiber being the Prym variety of an $N$-fold cover of $C$
\ref\sunref{R. Donagi, {\it Spectral covers}, Preprint 1994;
N. J. Hitchin, Duke Math. J. {\bf 54}  (1987) 91-114}.
Note that the $S$-duality getting mapped to $T$-duality of this
fiber torus is a very natural generalization of what
appears in the abelian case discussed above.  Moreover it suggests
an approach to showing $S$-duality for the non-abelian
four dimensional theory
by slicing the 4d path-integral in such a way that it becomes
equivalent to a family of abelian $S$-dualities glued
together in a nice way.

To get the precise form of the duality
we thus have to study the moduli space of a complex $d$ dimensional
torus. The moduli  space of a $2d$ real dimensional torus
is known \ref\narain{K. S. Narain, Phys. Lett B 169 (1986) 41 }\ to
be
$${SO(2d,2d)\over SO(2d)\times SO(2d)\times SO(2d,2d;{\bf Z})}$$
If we fix an integral K\"ahler form $k\in H^2(T^{2d}; {\bf Z})$ on
the torus
and ask about the moduli of complex structures
on the torus with that fixed K\"ahler class the answer is  described
as
follows\ref\math{P. Griffiths and J. Harris
{\it Principles of Algebraic Geometry} New York, Wiley, 1978}
:   Let $x^i,y^i, i=1,...,d$ denote
the coordinates of torus with periodicity 1 in each direction which
are chosen
so that
 the K\"ahler form can be written as
\eqn\intform{k=\sum_{i=1}^d n_i  dx^i \wedge dy^i}
where $n_i$ are positive integers.
Let $D$ denote the $d\times d$ diagonal matrix $D=(n_1,...,n_d)$.
Let  $z_i$
be the complex
coordinates of the torus.  Then we can choose them so that
\eqn\change{dz^i=n_i dx^i+\sum_j\Omega_{ij }dy^j}
where $\Omega$ is a  complex, symmetric $d\times d$ matrix with a
positive
definite
imaginary part  (all follow from the fact that $k$ defined above be a
positive
$(1,1)$ form) .
We have  $k=dz^i({1\over -2i Im \Omega})_{i\bar j}dz^{\bar j}$.  We
are
interested in how
the moduli space of complex structure and the particular
complexified K\"ahler
structure
(rescaling the fixed K\"ahler class by $t$
plus turning on an anti-symmetric $b$
field in the direction
of the fixed  K\"ahler form) imbed in the Narain moduli space.
There is an action of symplectic
group $Sp_J(2d)$ preserving the symplectic form
$$J=\left(\matrix{0&D\cr -D&0\cr}\right)$$
on the moduli of complex structure, and the full moduli space of
complex
structures is given by the quotient
$Sp_J(2d)/U(d)\times Sp_J(2d;{\bf Z})$, where $U(d)$ rotates the
$z_i$
among themselves.  Note that $Sp_J(2d)$
is equivalent  (and conjugate) to the canonical group
as far as they are defined over the reals, but the group
$Sp_J(2d;{\bf Z})$
very much
depends on $J$ (for example it would have been trivial if $n_i$ were
generic
real numbers).

We will now show that the relevant moduli space for our problem
 is split to the complex and K\"ahler directions,
where we just described the complex part.
Since the Narain moduli space is described as a group quotient, it
suffices to
show that
the generators of the complex deformations and the particular
K\"ahler deformation commute.  Let us first work over
the real numbers, in which case we can rescale coordinates
so that $D$ is replaced by the identity matrix and $J$ has
the canonical form.  It is
not difficult  to see that the generators of the deformations are
then given by
$${\rm Complex}: \qquad (\sigma_x\otimes S ;1\otimes A)/1\otimes A$$
$${\rm Kahler}:\qquad (t=\sigma_x\otimes 1;\ b=i\sigma_y\otimes J;\
\sigma_z\otimes J)
/\sigma_z\otimes J$$
where $S$ and $A$ denote symmetric and anti-symmetric generators of
$Sp(2d)$,
and the
Pauli matrices act on the $(L,R)$ decomposition of the Narain
momenta.
Note that the generators of K\"ahler deformations commute with those
of complex deformations and form the $Sp(2)$ (or $SL(2)$) group.  In
fact
this is the maximal subgroups of $SO(2d,2d)$ which commutes with
$Sp(2d)\subset
SO(2d,2d)$.
In order to find how the modular group acts on the $Sp(2)$, given
its imbedding in the Narain moduli, all we have to do is to find
integral
points
of the group generated by
$ \sigma_x\otimes 1,i\sigma_y\otimes J, \sigma_z\otimes J$;   We also
have to recall that we have rescaled coordinates
so that $J$ is in the canonical form.  If we undo this rescaling
and we
decompose
$J=\oplus J_i$ where $J_i$ corresponds to $i$-th direction
corresponding to $n_i$, we can view our $Sp(2)$ as sitting diagonally
in $\otimes Sp_i(2)$ where the common moduli $\tau$ is identified
as $n_i \tau_i$ in each subfactor.
With no loss of generality let us assume $n_i$'s have no common
divisor
(otherwise rescale the K\"ahler form so this is true). Let $n$ denote
the least common multiple of $n_i$.  Then it is clear that the common
intersection of all the $SL_i(2,Z)$ is generated by $T$ and $ST^nS$
where
$S:\tau \rightarrow -1/\tau$ and $T:\tau \rightarrow \tau+1$.  This
generates
the group $\Gamma_0(n)$.  We thus have the moduli space
$${Sp(2d)\over U(d)\times Sp_J(2d,{\bf Z})} \times { Sp(2)\over
U(1)\times
\Gamma_0(n)}$$
Thus the K\"ahler moduli of the Hitchin space has $\Gamma_0(n)$ as a
modular
group.
For $SU(N)$, all the $n_i$ are either $N $ or $1$, corresponding to
whether
they are related to combinations of cycles of $\hat C$ which are
projected to
trivial or non-trivial cycles of $C$.
 So in this case $n=N$ and we recover
the prediction of the $S$-duality that $\tau$ should belong to the
fundamental
domain
of $\Gamma_0(N)$.  In fact there is more information in the modular
transformation.
In particular  prediction of $S$-duality for
$\tau\rightarrow -1/\tau$ is in accord with the relation
between the Hitchin spaces for $SU(N)$ vs. $SU(N)/Z_N$.
\vskip 1cm

{\bf Acknowledgements}

We are grateful to T. Pantev for very patient discussions
on aspects of Hitchin spaces and to E. Witten for discussions
on 4d topological theories.
  C.V. would like to thank in
addition the hospitality of Rutgers University during the completion
of this work.

This research is supported in part by NSF grants PHY-92-18167 and
PHY-89-57162.  The research of M. B. was also supported by NSF 1994
NYI award
and DOE 1994 OJI award.
The research of C.V. was also supported by Packard fellowship.

\appendix{A}{Aspects of $N=2$ Reduction}

Let us consider some properties of the basis $\{ \alpha \}$ and the
connection
$E$.
Define a covariant derivative
$$\nabla_i=\partial/\partial X^i -iE_i~,~~
\nabla_{\bar{i}}=\partial/\partial X^{\bar{i}}-iE_{\bar{i}}~,$$
which acts on the space of Lie algebra valued functions
on ${\cal M}$.
By using eqs. \kahl, it is easy to check that $\partial_i
A_w=D_w E_i$ and
$\partial_{\bar{k}} A_{\bar{w}}=D_{\bar{w}} E_{\bar{k}}$, i.e.
$[\nabla_i
,D_w]=0=
[\nabla_{\bar{i }},D_{\bar{w}}]$.
The only non-zero component of the curvature is $(1,1)$
$$\Phi_{i \bar j}=i[\nabla_i, \nabla_{\bar j}]~,$$
so that $\nabla$ is holomorphic. It is also easy to check that
$\nabla_i
\alpha_{w\bar{k}}
=D_w \Phi_{i\bar{k}}$,
$\nabla_{\bar{k}}\alpha_{{\bar{w}}i}=-D_{{\bar{w}}}\Phi_{i\bar{k}}$
and
$D_i\alpha_{{\bar{w}}j}=D_j\alpha_{{\bar{w}}i}$,
$D_{\bar{i}}\alpha_{w\bar{j}}=D_{\bar{j}}\alpha_{w\bar{i}}
$.
The Christoffel connections $\Gamma^k_{ij}$ and
$\Gamma^{\bar{k}}_{\bar{i}\bar{j}}$ can be constructed in terms of
the basis
$\alpha_{{\bar{w}}i}$, $\alpha_{w\bar{k}}$ as follows
\eqn\Chr{\Gamma_{ij,\bar{k}}= \int_C {\rm Tr}~\alpha_{w\bar{k}}
\nabla_j\alpha_{{\bar{w}}i} = \partial_j G_{i\bar{k}},~~~
\Gamma_{\bar{i}\bar{j},k}= \int_C {\rm
Tr}~\alpha_{{\bar{w}}k}\nabla_{\bar{i}}
\alpha_{w\bar{j}} =\partial_{\bar{i}} G_{k\bar{j}}.}
Notice that the other components vanish since the metric is K\"ahler.
The basis vectors  $\alpha_{\bar{w}i}$ ($\alpha_{w\bar{k}}$) are
covariantly
constant with respect to the covariant derivatives $\nabla_i
\delta^k_j -
\Gamma^k_{ij}$ ($\nabla_{\bar{i}} \delta^{\bar{k}}_{\bar{j}} -
\Gamma^{\bar{k}}_{\bar{i}\bar{j}}$) which act on  Lie algebra
valued 1-forms on ${\cal M}$, i.e. $\nabla_i \alpha_{\bar{w}j}=
\Gamma^k_{ij}\alpha_{\bar{w}k}$ and $\nabla_{\bar{i}}
\alpha_{w\bar{j}}=
\Gamma^{\bar{k}}_{\bar{i}\bar{j}}\alpha_{w\bar{k}}$.
The latter equations follow from the fact that $D_{w}(\nabla_i
\alpha_{{\bar{w}}j})=0=D_{\bar{w}}(\nabla_{\bar{k}}\alpha_{w\bar{j}})$
{}.

The Riemann tensor can also be  written down in terms of the basis
vectors
as follows
\eqn\Riemann{R_{i\bar{k}j\bar{l}}=\int_C(D_i\alpha_{w\bar{k}}~
D_{\bar{l}}\alpha_{\bar{w}j}+D_j\alpha_{w\bar{k}}~
D_{\bar{l}}\alpha_{\bar{w}i}).}

For convenience we also discuss briefly the equations of motion for
$A_{\S}$
and $\phi$.
These equations of motion read
\eqn\Asig{D_w F_{\bar{z}\bar{w}}= -i \{ \lambda_{\bar{z}\bar{w}},
\chi_w \},
{}~~~
D_{\bar{w}} F_{zw}=i \{ \lambda_{zw}, \chi_{\bar{w}} \},}
and
\eqn\ephi{D_w D_{\bar{w}} \phi = i\{\chi_w,\chi_{\bar{w}} \}, ~~~~
D_{\bar{w}} D_w \bar{\phi} =
ig^{z\bar{z}}\{\lambda_{zw},\lambda_{\bar{z}\bar{w}} \}.}
We assume that the connection $A_C$ is irreducible and therefore the
solutions of equations \Asig\ and \ephi\ are unique.
At first glance the solutions to these equations are expected to be
non-local.
However, by using an identity $[\alpha_{\bar{w}i},
\alpha_{w\bar{k}}]=iD_{\bar{w}}
D_{w}
\Phi_{i\bar{k}}$ one can reduce the solutions to a local form.
In particular for $\phi$ and $\bar{\phi}$ we get eq.\exphi .

\appendix{B}{Aspects of $N=4$ Reduction}

The cotangent vectors $\delta A_C$, $\delta \phi_{w}$ and
$\delta\phi_{\bar{w}}$ to the moduli space ${\cal M}^H$
obey the equations which are variation of
eq.\hitreduce .
To study the moduli space of the Hitchin's equations it is convenient
to choose a special gauge
\eqn\hitgauge{D^w \delta A_w +D^{\bar{w}}\delta A_{\bar{w}}+ i[\delta
\phi_{\bar{w}},\phi_{w}]+i[\phi_{\bar{w}},\delta\phi_{w}]=0.}
Thus the cotangent vectors obey the following equations
\eqn\hitcotan{D^w \delta A_w= -i[\delta \phi_{\bar{w}},\phi_{w}],~~~
D^{\bar{w}}\delta\phi_{\bar{w}}=i[\delta A_w,\phi_{\bar{w}}].}

To introduce explicitly the collective (real) coordinates $\{X^A\}$
it
is
convenient to choose a basis $\{(\alpha_{A\bar{w}},\beta_{A
w})\}$ and
$\{(\alpha_{A  w},\beta_{A \bar{w}})\}$ on the space of pairs
$(\delta A_w
,\delta\phi_{\bar{w}})$ and $(\delta A_{\bar{w}},\delta\phi_w)$.
Then the variation $(\delta A_w ,\delta\phi_{\bar{w}})$, $(\delta
A_{\bar{w}},\delta\phi_w)$ can be decomposed along the basis
modulo a gauge transformation as follows
\eqn\basisP{\partial_{A} A_w=\alpha_{A w} +D_w E_{A},~~~
\partial_{A} A_{\bar{w}}=\alpha_{A \bar{w}}+D_{\bar{w}} E_{A},}
$$\partial_{A}\phi_{w}=\beta_{A w}-i[\phi_{w},
 E_{A}],~~~\partial_{A}\phi_{\bar{w}}=\beta_{A\bar{w}}-i[\phi_{
\bar{w}}, E_{A}],$$
where $\partial_{A}=\partial /\partial X^{A}$, and $E_{A}$ can
be identified with a connection on ${\cal M}^H$.
The moduli space  ${\cal M}^H$ is hyperK\"ahler and has a natural
hyperK\"ahler metric
\eqn\HyperM{G_{AB}=\int_C {\rm Tr}(\alpha_{A
w}\alpha_{B\bar{w}}+\beta_{A w}\beta_{B\bar{w}}+(A
\leftrightarrow B)), }
induced by a bilinear form on the space of pairs $(\delta A_{\alpha}
,\delta\Phi_{\alpha})$
($\alpha=1,2$ are Lorentz indices on $C^h$)
\eqn\hitmetric{g((\delta A ,\delta \Phi),(\delta A,\delta\Phi))=
\int_C {\rm Tr} (\delta A \wedge \ast \delta A+\delta\Phi  \wedge
\ast \delta\Phi),}
where and $\phi_w=\Phi_1+i\Phi_2$, $\phi_{\bar{w}}=\Phi_1-i\Phi_2$.
Similarly we introduce a K\"ahler form
\eqn\HyperO{\Omega^I_{AB}=\int_C {\rm Tr}(\alpha_{A
w}\alpha_{B\bar{w}}+\beta_{A w}\beta_{B\bar{w}}-(A
\leftrightarrow B)),}
on ${\cal M}^H$ induced by a symplectic form
\eqn\hitkahler{\omega^I ((\delta_1 A ,\delta_1 \Phi),(\delta_2
A,\delta_2 \Phi))=
\int_C {\rm Tr} (\delta_1 A \wedge \delta_2 A+\delta_1\Phi  \wedge
\delta_2\Phi).}
It is easy to check that this form is closed.
We naturally define the complex structure
$I^{A}_{B}=G^{AC}\Omega^I_{BC}$
on  ${\cal M}^H$ .

With a complex structure $I$ one can choose the complex coordinates
$X^{\mu}$
and  $X^{\bar{\mu}}$ on ${\cal M}^H$.
It is easy to check that the only non-vanishing components of the
basis vectors
are
$(\alpha_{\mu \bar{w}},\beta_{\mu w})$ and $(\alpha_{\bar{\mu}
w},\beta_{\bar{\mu} \bar{w}})$.

Since the moduli space is hyperK\"ahler there are two more complex
structures
$J$ and $K$ which satisfy the algebraic identities for the
quaternions.
One of these complex structures, $J$, can be defined as
$J^{A}_{B}=G^{AC}\Omega^J_{BC}$, where
$\Omega^J_{BC}$ is
a symplectic form on ${\cal M}^H$, which is induced by a symplectic
form
\eqn\complexJ{\omega^J  ((\delta_1 A ,\delta_1\Phi),(\delta_2
A,\delta_2\Phi))=\int_C {\rm Tr} (\delta_1 A \wedge \delta_2\Phi
-\delta_1\Phi \wedge \delta_2 A) .}
The symplectic form $\Omega^J$ has only non-vanishing components
\eqn\OmJ{\Omega^J_{\mu\nu} =\int_C {\rm Tr}(\alpha_{\mu
\bar{w}}\beta_{\nu w}-\alpha_{\nu
\bar{w}}\beta_{\mu w}) ~~~{\rm and}~~~
\Omega^J_{\bar{\mu}\bar{\nu}} =\int_C {\rm Tr}(\alpha_{\bar{\mu}
w}\beta_{\bar{\nu} \bar{w}}-\alpha_{\bar{\nu}w}\beta_{\bar{\mu}
\bar{w}}),}
which are anti-holomorphic and holomorphic on the moduli
space ${\cal M}^H$, respectively, and obey the following equation
\eqn\JG{\Omega^J_{\mu\nu}G^{\nu\bar{\nu}}
\Omega^J_{\bar{\mu}\bar{\nu}}
=G_{\mu\bar{\mu}}.}

Similar to the analysis of the moduli space of flat connections
we see that
$[\nabla_{\mu}, D_w] =0$, $[\nabla_{\bar{\mu}}, D_{\bar{w}}] =0$,
$[\nabla_{\mu},\phi_{\bar{w}}]=0$ and
$[\nabla_{\bar{\mu}},\phi_w]=0$, where
 $\nabla_{\mu}=\partial_{\mu}-i E_{\mu}$ and
$\nabla_{\bar{\mu}}=\partial_{\bar{\mu}}-i E_{\bar{\mu}}$.
One can also easily check that $[\nabla_{\mu},\nabla_{\nu}]=0$ and
$[\nabla_{\bar{\mu}},\nabla_{\bar{\nu}}]=0$, and hence $\nabla_{\mu}
\alpha_{\bar{w}\nu}$ ($\nabla_{\bar{\mu}}\alpha_{w\bar{\nu}}$) and
$\nabla_{\mu}\phi_{w\nu}$
($\nabla_{\bar{\mu}}\phi_{\bar{w}\bar{\nu}}$) are
symmetric with respect to indices $\mu ,~\nu$ ($\bar{\mu}
,~\bar{\nu}$).
This follows from the fact that the forms on ${\cal M}^H$ $\Phi_{\mu
\nu}=i[\nabla_{\mu},\nabla_{\nu}]$ and
$\Phi_{\bar{\mu}\bar{\nu}}=i[\nabla_{\bar{\mu}},\nabla_{\bar{\nu}}]$
are
annihilated by
the operator $D_{\bar{w}}D_w-[\phi_w ,[\phi_{\bar{w}} ,\cdot]]$ (we
assume that
a non-trivial magnetic flux through $C^h$ is turned on).

It is also worth noticing that the complex structure $J$ exchanges
the $\alpha$
and $\beta$ components of the basis as follows
\eqn\exch{J^{\mu}_{\bar{\mu}} \alpha_{\bar{w}\mu}=
\beta_{\bar{w}\bar{\mu}},
{}~~~J^{\mu}_{\bar{\mu}} \beta_{w\mu}=
-\alpha_{w\bar{\mu}}, ~~~
J^{\bar{\mu}}_{\mu} \alpha_{w\bar{\mu}}= \beta_{w\mu},
{}~~~
J^{\bar{\mu}}_{\mu} \beta_{w\bar{\mu}}= -\alpha_{w\mu}.}
As a generalization of the $N=2$ case we also have
\eqn\covHit{\nabla_{\nu}
\alpha_{\bar{w}\mu}=\Gamma^{\lambda}_{\mu\nu}
\alpha_{\bar{w}\lambda}- iJ^{\bar{\mu}}_{\mu}
[\Phi_{\nu\bar{\mu}},\phi_{\bar{w}}],
{}~~~
\nabla_{\nu}\beta_{w\mu}=\Gamma^{\lambda}_{\mu\nu}\beta_{w
\lambda}-
J^{\bar{\mu}}_{\mu}D_{w}\Phi_{\nu\bar{\mu}},}
and similar relations for $\alpha_{w\bar{\mu}}$ and
$\beta_{\bar{w}\bar{\mu}}$.
Here $\Phi_{\mu \bar{\mu}}=i[\nabla_{\mu},\nabla_{\bar{\mu}}]$, and
$\Gamma^{\lambda}_{\mu\nu}$ stand for the Christoffel connection
which is
defined as follows
\eqn\ChrHit{\Gamma_{\mu\nu,\bar{\lambda}} = \int_C
(\alpha_{w\bar{\lambda}} \nabla_{\mu}
\alpha_{\bar{w}\nu}+\beta_{\bar{w}\bar{\lambda}}
\nabla_{\mu}\beta_{w\nu})=\partial_{\mu}
G_{\nu\bar{\lambda}},}
$$\Gamma_{\bar{\mu}\bar{\nu},\lambda} = \int_C
(\alpha_{\bar{w}\lambda}
\nabla_{\bar{\mu}}
\alpha_{w\bar{\nu}}+\beta_{w\lambda}
\nabla_{\bar{\mu}}\beta_{\bar{w}\bar{\nu}})=\partial_{\mu}
G_{\lambda\bar{\nu}}.$$
One may wonder if eqs.\covHit \ are consistent with the symmetry of
the
Christoffel connection.
It is easy to see that the consistency condition reads
$J^{\bar{\mu}}_{\nu} \Phi_{\mu\bar{\mu}}J_{\bar{\nu}}^{\mu}
=-\Phi_{\nu\bar{\nu}}$.

Let us  continue our discussion of the reduction for $N=4$ in the
case of
partially twisted
model (in $C$-direction).
By substituting the solutions to the Hitchin's equations for $A_C$,
$\phi_w$
and $\phi_{\bar{w }}$ into the action one can see that in the limit
$\epsilon\to 0$ the rest of the bosonic fields enter quadratically
into the action.
They are scalar on $C^h$ and therefore do not correspond to any
propagating
degrees of freedom in the effective 2D $\sigma$-model provided the
fields
$A_C$, $\phi_{w}$ and $\phi_{\bar{w}}$ correspond to irreducible
configuration
(for example, one can consider the $SO(3)$ gauge group with a
non-trivial flux through $C^h$).
By the equations of motion these bosonic fields are reduced
to certain combinations of the fermionic fields similar to the
$N=2$ case.
The solutions of the equations of motion for
the scalar fields (on $C^h$) can be expressed in local form similar
to those of the $N=2$ case
due to the following equation
\eqn\Phit{D_{\bar{w}}D_w \Phi_{\mu \bar{\mu}}-[\phi_w
,[\phi_{\bar{w}}
,\Phi_{\mu \bar{\mu}}] ]=
i[\alpha_{w\bar{\mu}}, \alpha_{\bar{w}\mu}]+
i[\beta_{\bar{w}\bar{\mu}},
\beta_{w\mu}].}

Integrating out the fermionic fields which are
scalars on $C^h$ one gets
constraints on the fermionic fields which have a vector index on
$C^h$.
By these constraints the latter fields are naturally split into pairs
$(\chi^a_{w+},\bar{\lambda}^a_{\bar{w}-})$,
$(\chi^a_{\bar{w}+},\lambda^a_{w-})$ which are enforced  to obey the
eqs.\hitcotan\ for cotangent vectors on the moduli space ${\cal
M}^H$, and
hence can be decomposed along the above chosen basis.
Thus these fields give rise to 2d fermionic fields $\psi^{\mu}_+$,
$\psi^{\bar{\mu}}_+$,
$\psi^{\mu}_-$ and $\psi^{\bar{\mu}}_-$.
Finally for the dimensionally reduced theory we get the standard
action for a supersymmetric 2d sigma-model
\eqn\Hitaction{\eqalign{S^H= {1\over e^2}\int_\S {\rm d}^2z~
G_{\mu\bar{\mu}}(&{1\over
2}\partial_z X^{\mu}~\bar{\partial}_{\bar{z}} X^{\bar{\mu}} +{1\over
2}
\bar{\partial}_{\bar{z}} X^{\mu}~\partial_z X^{\bar{\mu}}+\cr
&\psi^{\mu}_+ \overline{D}_{\bar{z}}\psi^{\bar{\mu}}_+ +
\psi^{\mu}_- D_z\psi^{\bar{\mu}}_-)
 - R_{{\mu}\bar{\mu} {\nu}\bar{\nu}} \psi^{\bar{\mu}}_+ \psi^{\mu}_-
\chi^{\bar{\nu}}_- \psi^{\mu}_+.}}
Here the covariant derivatives $D_z$ and $\overline{D}_{\bar{z}}$ are
constructed by pulling back
the Christoffel connection on the tangent bundle $T{\cal M^H}$ to
${\cal M^H}$.

The contributions from the path integral over quadratic fluctuations
orthogonal to the zero modes cancel due to supersymmetry.
Notice also that the fermion number current is non-anomalous
similar to the one which appears in unreduced 4d $N=4$ supersymmetric
Yang-Mills theory.

We can also consider the totally twisted 4d N=4 Yang-Mills theory.
By the dimensional reduction
(with the twist used in ref.\Sdual) we get the Lagrangian of a
twisted version
of the
above supersymmetric $\sigma$-model.
The twisting current  has a bosonic piece and hence some of the
bosonic fields become 1-forms on the world sheet $\S$.
In $4$-dimensional theory this current generates $U(1)$
global phase rotations of $\phi$ and $\bar{\phi}$.
In terms of the {\it partially} twisted theory the bosonic
contributions to the
current is
$j_n = {\rm Tr}( \bar{\phi}^w~\partial_n \phi_w
-\partial_n\bar{\phi}^w ~\phi_w
+...)$,
where $n$ is a worldsheet index on $\S$.
Under the dimensional reduction this current becomes
\eqn\btwist{j_n=\int_C   {\rm Tr}( \bar{\phi}_{\bar{w}}(X) \partial_n
{\phi}_{\bar{w}}(X) - \partial_n {\bar{\phi}}_{\bar{w}}(X)\phi_w (X))
+({\rm
fermionic ~~terms}),}
where $X(z,\bar{z})$ determines the map $\S \to {\cal M^H}$.
The fields $\bar{\phi}_{\bar{w}}$ and $\phi_w$ obey the Hitchin's
equations
and hence are functions on ${\cal M}^H$.
  In fact, in the $\sigma$-model $j_n$ is a Noether current
corresponding to
the  action of  $U(1)$ on ${\cal M}^H$ given by $(A,\phi)\to (A,
e^{i\theta}\phi)$.  This group action is Poisson with respect to the
symplectic
form $\omega^I$ \hitkahler\ and the hamiltonian $\mu =\int_C {\rm
Tr}\,\bar{\phi}_{\bar{w}}\phi_w$.
By the equations of motion of the $\sigma$-model $j_n$ is conserved.

Actually the above splitting of the coordinates on ${\cal M}^H$
naturally
appears
under the dimensional reduction of the totally twisted 4d N=4 SYM
theory.
In this case we start with Hitchin's fields $\phi_{zw}$ and
$\phi_{\bar{z}\bar{w}}$ which are
2-forms on $M$.
Under the dimensional reduction we have to assign the worldsheet
indices
$z,\bar{z}$ to some of
the components of the basis vectors, and hence to some of collective
coordinates.

To understand better this twisting procedure recall that ${\cal M}^H$
is
fibered over ${\bf C}^{3g-3}$
with ${\rm det} \phi_w$ projecting onto the base \hit. On the base
${\bf
C}^{3g-3}$ let us introduce the standard affine coordinates $Y^1,
\cdots,
Y^{3g-3}$, which constitute half of $6g-6$ coordinates on the full
space ${\cal
M}^H$. In such a description the $U(1)$ generated by \btwist\ acts by
the
phase rotation $Y^I \to e^{i \theta } Y^I$. Let us denote by $Z^i$
the
coordinates along
fiber, which are inert under the $U(1)$ rotations. In this
coordinate system the
action looks as
follows
\eqn\Hitact{\eqalign{S^H &= {1 \over 2e^2} \int_\S {\rm d}^2z~\Big(
G_{i\bar{k}}(
 \partial_z Z^i~\bar{\partial}_{\bar{z}} Z^{\bar{k}}
+\bar{\partial}_{\bar{z}}
Z^i ~\partial_z Z^{\bar{k}} )+\cr
&G_{I\bar{I}} g^{z \bar z}( D_z Y^I_z ~\overline{D}_{\bar{z}}
Y^{\bar{I}}_{\bar{z}} + \overline{D}_{\bar{z}} Y^I_z~D_z
Y^{\bar{I}}_{\bar{z}}
)+\cr
& C_{i\bar{I},J}     Y^J_z g^{z \bar z}
( \bar{\partial}_{\bar{z}} Z^i~D_z Y^{\bar{I}}_{\bar{z}}+ \partial_z
Z^i ~\overline{D}_{\bar{z}}  Y^{\bar{I}}_{\bar{z}})+\cr
&C_{I\bar{k},\bar{I}}  Y_{\bar{z}}^{\bar{I}} g^{z \bar z}
( \overline{D}_{\bar{z}} Y^I_z~\partial_z Z^{\bar{k}}+
D_z Y^I_z~\bar{\partial}_{\bar{z}} Z^{\bar{k}} )+\cr
&~~~~~~~~~~{\rm fermion~terms} \Big)~,\cr}}
where $C_{i\bar{I},J}Y^J=G_{i\bar{I}}$, $C_{I\bar{k},\bar{I}}Y^{\bar
I}=G_{I\bar{k}}$.
All the terms in the action
have equal number of
$Y^I_z$ and $Y^{\bar I}_{\bar{z}}$ which are contracted
by appropriate number of $g^{z\bar z}$. This is the action for a
new
topological $\sigma$-model,
which is nothing but  $N=4$ superconformal theory twisted
with a current discussed above.  This theory
computes the Euler class of moduli of holomorphic maps to Hitchin
space, as is clear from similar constructions in the literature
\ref\mq{V. Mathai and D. Quillen, ``Superconnections, Thom Classes,
and Equivariant Differential Forms,'' Topology {\bf 25} (1986)
85.}\ref\atj{M.
F. Atiyah and L. Jeffrey, ``Topological Lagrangians And
Cohomology,'' J. Geom. Phys. {\bf 7} (1990) 119.}\ref\bs{L. Baulieu
and I. M.
Singer, ``Topological Yang-Mills
Symmetry,''
Comm. Math. Phys. {\bf 117} (1988) 253.}\ref\ugmit{E.
Witten,``Introduction To
Cohomological Field
Theories,''
Int. J. Mod. Phys. {\bf A6} (1991) 2775.}\ref\oldwit{E. Witten, ``The
$N$
Matrix Model And Gauged WZW
Models,''
Nucl. Phys. {\bf B371} (1992), 191}\ref\distler{J. Distler, ``Notes
On $N=2$
Sigma Models,''
hep-th-9212062.}\ref\cgm{S. Cordes, G. Moore, and S. Ramgoolam,
``Large $N$  2D
Yang-Mills Theory \& Topological String Theory,''
hepth-9402107}\ref\thbl{M.
Blau and G. Thompson,
Comm. Math. Phys. {\bf 152} (1993) 41;
Int. J. Mod. Phys. {\bf A 8} (1993) 573 }\Sdual .
In this context using the results of \Sdual , we have a prediction
for the Euler characteristic of moduli of holomorphic maps
from $\Sigma$ to ${\cal M}^H(C)$.
A similar construction
has been considered in the context of $N=2$ string with cotangent
of a Riemann surface as the target \ref\mooret{
R. Dijkgraaf and G. Moore, private
communication.}.

\appendix{C}{}
 Here we would like to sketch the derivation of  formulas   \tor,
\hig\
computing the correlation function $\langle e^{\alpha a +\beta b}
\rangle ^{M^4
 }_z$ of $N=2$ topological supersymmetric Yang Mills theory in a
sector with
fixed `t Hooft flux $z$.  To do it, one can use either  the ``cosmic
strings''
\WiI\ or the Monopole Equation \WiII. The four-manifold $M^4  $ for
us is
either $T^2\times C^h$ or $\S \times C^h$, where both genera  $g$ and
$h$
satisfy $g,h>1$. For the case  $M^4  =T^2\times C^h$ the  canonical
divisor
consists of  $2g-2$ {\it nonintersecting} components (``cosmic
strings'').
When $M^4  =\S \times C^h$  the cosmic string is a  Riemann surface
of genus
$8(g-1)(h-1)+1$.  Indeed, the canonical class of  $\S \times C^h$ is
very ample
so  one can choose its divisor to be a smooth connected curve. It
follows from
Bertini's theorem \ref\Hart{R.~Hartshorn, {\it Algebraic Geometry},
Springer,
1977} Ch.~2, Prop.~8.18 and the standard technical lemma \Hart\
Ch.~3,
Prop.7.9.
Then the formula (2.79) from \WiI\ gives \tor\ and \hig\  (we need to
fix the
right normalization though, see the footnote after \tor).

The other way to obtain \tor\ and \hig\ is through the Monopole
Equation \WiII.
 The arguments in Section 4 there tell us that the basic class $x \in
H^2(\S
\times C^h)$ should satisfy the following conditions:
1) It comes from $H^2(\S ,{\bf Z})+H^2( C^h,{\bf Z})$:  $x=x(\S
)\omega
_\Sigma+x(C^h)\omega _C$ and  is divisible by 2: $x(\S )\equiv 0({\rm
mod} 2)$,
$x(C^h)\equiv 0({\rm mod} 2)$;
2)  $x^2=4(g-1)(h-1)$;
3) $1-g \leq x(\S )/2\leq g-1$ and $1-h \leq x(C^h)/2\leq h-1$. These
constraints fix $x=\pm (2(g-1)\omega _\Sigma+2(h-1)\omega _C)$ to be
(plus or
minus) the canonical class $K$.  The correlation function is given by
\WiII\
\eqn\MonEq{\eqalign{
\langle e^{\alpha a +\beta b} \rangle ^{\S \times C^h}_z= {1\over
2}\sum_{x=\pm
K}(-1)^{z\cdot x/2} n_x (e^{\alpha t x(C)+\beta \lambda}+i^{z^2}
e^{-i\alpha t
x(C)-\beta \lambda})
}}
where $t$ is the normalization constant and $n_x$ is
``multiplicity''.  The
absolute value of  $n_x$ computes the number of  decompositions
$\alpha \beta
=\eta$ where $\eta$ is a fixed holomorphic 2-differential --- a
section of $K$,
 $\alpha$ is a section of the line bundle ${\cal L}\otimes K^{1\over
2}$ and
$\beta$ is a section of  the line bundle ${\cal L}^{-1}\otimes
K^{1\over 2}$.
The bundle ${\cal L}$ should satisfy $2c_1({\cal L})=x$. Since $x=\pm
K$,
either ${\cal L}=K^{1\over 2}$ and then $\alpha =\eta$, $\beta=1$ or
${\cal
L}=K^{-{1\over 2}}$ and  $\alpha =1$, $\beta=\eta$. In both cases,
$|n_x|=1$.
The sign of  $n_x$  can be obtained by the cohomological computation
suggested
in \WiII\ which gives $n_K=1$ and $n_{-K}=(-1)^{(g-1)(h-1)}$  which
finally
leads to \hig\ in the main text.

Using the Monopole Equation one can also  verify the results of
\WiI\ for
$M^4  =T^2\times C^h$.
The same constraints 1)--3) single out the basic classes
$\{x=2j\omega_C\,|
1-h\leq j\leq h-1\}$. The correlation function  $\langle e^{\alpha a
+\beta b}
\rangle ^{T^2\times C^h}_z$    is computed by the  equation like
\MonEq,  only
now one has to sum over all $x$ from the list above.  Also in this
case  the
line bundle ${\cal L}$ comes only from  the curve $C$. So $|n_x|$
computes the
number of ways to split  1-differential $\eta$ {\it on} $C$ as a
product of two
sections. For ${\rm deg}{\cal L}=j$,  the section $\alpha$ has
$j+h-1$ zeroes
and $\beta$ has $h-1-j$ zeroes. Together they form $2h-2$ zeroes of
$\eta$.
Since the divisor of zeroes determine the line bundle on $C$
completely,  {\it
each} splitting of  zeroes of $\eta$ into two groups of $j+h-1$ and
$h-1-j$
elements gives a solution so $|n_x|=C^{2h-2}_{j+h-1}$.  Also, the
computation
gives ${\rm sgn}\,(n_x)=(-1)^{j+h-1}$ and we arrive at \tor.
\listrefs
\end